\documentclass[final,5p,times]{elsarticle}

\usepackage{amssymb}
\usepackage{amsmath}

\usepackage{hyperref}
\journal{Physics Letters A}

\usepackage[dvipsnames]{xcolor}
\usepackage{xspace}
\usepackage{bm}
\usepackage{siunitx}
\usepackage{upgreek}

\newcommand\md{\ensuremath{\mathrm{d}}\xspace}
\newcommand\Ai{\ensuremath{\mathrm{Ai}}\xspace}
\newcommand\Hbar{\ensuremath{\overline {\mathrm{H}}}\xspace}
\newcommand\gbar{\ensuremath{g}\xspace}

\begin{document}

\begin{frontmatter}



\title{Single-bounce quantum gravimeter to measure the free-fall of anti-hydrogen}

\affiliation[LKB]{organization={Laboratoire Kastler Brossel,Sorbonne Universit{\'{e}}, CNRS, ENS-Universit{\'{e}} PSL, Coll{\`{e}}ge de France},
            addressline={4 place Jussieu},
            city={Paris},
            postcode={75005},
            state={},
            country={France}}

\author[LKB]{Joachim Guyomard} 
\author[LKB]{ Pierre Clad\'e}
\author[LKB]{Serge Reynaud}

\begin{abstract}
We propose an innovative concept for a quantum gravimeter, where atoms prepared in a Heisenberg-limited state perform a single bounce on a mirror followed by a free fall. This quantum gravimeter produces a simple and robust interference pattern which should allow to measure the free-fall acceleration of atoms. We estimate the expected accuracy of the measurement in the GBAR experiment, which aims at testing the equivalence principle on anti-hydrogen at CERN antimatter facilities. Using simulations and estimation techniques based on Cramer-Rao law and Fisher information, we show that the new quantum sensor improves the expected accuracy of the measurement. The proposal opens the door to free fall measurements on rare or exotic atomic species, especially in situations where experimental time or detection events are limited by intrinsic physical reasons.
\end{abstract}



\begin{keyword}



 
\end{keyword}

\end{frontmatter}

As the asymmetry between matter and antimatter in the Universe is one of the fundamental open questions in modern physics, testing the Equivalence Principle on anti-matter remains a key challenge for which experimental knowledge is now available \cite{Anderson2023}. 
The Weak Equivalence Principle \cite{Will2014llr}, the universality of free fall independently of the nature and mass of the probe, has been tested with high accuracy on macroscopic matter bodies \cite{Wagner2012,Viswanathan2018,Touboul2022prl,Touboul2022cqg} as well as on atoms \cite{Kajari2010apb,Herrmann2012cqg,Barrett2015njp,Asenbaum2020,Tino2020ppmp}. Ambitious projects are developed at CERN antimatter facilities to measure the free fall acceleration $\gbar$ of antihydrogen ($\Hbar$) atoms in Earth gravity field  \cite{Huber2000asr,Bertsche2015,bertsche2018,Yamazaki2020}.
Among them, the GBAR experiment aims at  measuring ${\gbar}$ at the $1\%$ level by timing the classical free fall of ultra-cold antihydrogen atoms \cite{Indelicato2014,Perez2015,Adrich2023}.

In this letter we propose to improve this expected accuracy by several orders of magnitude by using a quantum interference measurement of the free fall of the atoms rather than timing their classical free fall. The main idea is to let freely falling atoms bounce on a mirror surface positioned on their trajectory, due to quantum reflection on the Casimir-Polder interaction when they approach the surface \cite{Shimizu2001,Druzhinina2003,Oberst2005,Pasquini2006}. 
The Casimir-Polder interaction is effective on $\Hbar$ as well as on atoms and it should produce quantum reflection and prevent their annihilations on the surface \cite{Voronin2005jpb,Dufour2013,Crepin2017epl}.
Gravity and reflection form a trap and produce bound quantum states, the so-called  Gravitational Quantum States (GQS) which have been observed on ultracold neutrons \cite{Nesvizhevsky2002nature,Nesvizhevsky2009}, for which reflection is produced by repulsive Fermi potential.

In a previous proposal \cite{Crepin2019pra} the antihydrogen atoms were supposed to undergo a large number of bounces and thus spend a long time in bound states so that they acquired large phase-shifts for amplitudes corresponding to different states. This implied that the interference pattern had a complex structure (see the Figure 5 in \cite{Rousselle2022epjd}), which contained the information needed to extract the value of ${\gbar}$. This complexity appears to be a source of worries however, since the interference pattern is sensitive to many details to be accounted for in the theoretical calculation of the signal or its experimental measurement. 

The new concept proposed in this letter is a quantum interferometry measurement of free fall, aimed at solving this complexity problem.
Quantum interferences are now produced for freely falling atoms after a single bounce on the mirror. The interference pattern has the simple structure discussed below, and leads to a much more robust comparison of theory and experiment. The scheme is completely different from matter-wave interferometers commonly used to measure free fall, where interferences are produced by superpositions of waves having followed different trajectories after beam splitters \cite{kasevich1991,Storey1994,Borde2001,Gillot2014}. We will show that the new scheme produces an interferometric measurement of ${\gbar}$ with a good accuracy, even with a limited sample of detected events. 

We will illustrate the application of the new concept by applying it to the measurement of ${\gbar}$ in the GBAR experiment. The expected accuracy will be evaluated quantitatively by Monte-Carlo simulations compared to optimal estimation techniques based on Cramer-Rao law and Fisher information.
With these techniques, it will be shown to be improved by 4 orders of magnitude in comparison to the classical experiment. The expected accuracy will be shown to be even better in the single-bounce scheme than that already obtained with many bounces. All statements and numbers are given for the same resource, that is the same number of $\Hbar$ atoms prepared in the same ultracold quantum state.
 
\begin{figure}[h]
\begin{center}    
    \includegraphics[width=.8\linewidth]{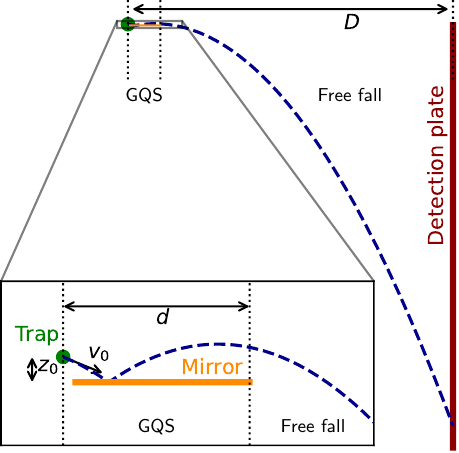}
\caption{Schematic representation of the experimental setup. A Gaussian wave-packet prepared in a trap falls down onto a mirror where it experiences a single bounce. Subsequent free fall reveals interferences which are recorded as positions of atoms on a detection plate. } 
\label{fig:onebouncescheme}
\end{center}
\end{figure}

The single bounce quantum gravimeter is schematized on Fig.\ref{fig:onebouncescheme}. The initial state is a Gaussian wave-packet prepared at Heisenberg limit. In GBAR,  ${\Hbar}^+$ ions are prepared in the ground state of an ion trap and ${\Hbar}$ atoms are released by photo-detaching the excess positron \cite{walz2004grg}. The laser pulse used for the photo-detachment is the Start signal of a time-of-flight measurement with the Stop signal given by annihilation at detection plate. 
As the horizontal distance $D$ from the source to the detection plate is fixed (and the size of the source negligible compared to $D$), the value of horizontal velocity $V$, preserved during the whole flight, is precisely deduced from the time-of-flight for each detection event. In the following, we measure 
time of evolution $t$ as the run horizontal distance $x=V\,t$ and use a standard value $V_0=1$~m~s$^{-1}$ of $V$ for fixing numbers and plots. 

The important parameters for the initial Gaussian wave-function are the mean altitude $z_0$ measured above the mirror plate (lying at $z=0$), the mean vertical velocity $v_0$ and the altitude dispersion $\sigma_z$. The latter is fixed by the harmonic trap frequency $f_0$ and determines velocity dispersion through the Heisenberg relation $\sigma_v=\frac\hbar{2m\sigma_z}$. The numbers used below are $z_0 = 1$~mm, $v_0= -{91.5}$~mm~s$^{-1}$, and 
$\sigma_z=0.4$~$\mu$m which corresponds to numbers matching the GBAR experiment with $f_0=30$~kHz \cite{Hilico2014} and $m\simeq1$ atomic mass unit.
 
The first phase of evolution of the wave-function corresponds to left-hand side on Figure \ref{fig:onebouncescheme}. It starts at initial time $t_0=0$ and includes the bounce on the mirror and the free fall before and after it, till the end of the mirror at $x=d$. This evolution is decomposed over the GQS which solve Schrödinger equation in presence of gravity and reflection (assumed perfect at this stage)
\begin{equation}
\psi_{t}(z)=\sum_n c_n\,\chi_n(z) \,e^{ -i \frac{E_n\, t}{\hbar}} 
\quad,\quad 0<Vt<d \,.
\label{eq:discrete_decomposition}
\end{equation}
The eigen-functions $\chi_n(z)$ are Airy functions corresponding to  energies $E_n$. The coefficients $c_n$ are defined so that the expression \eqref{eq:discrete_decomposition} reproduces the initial Gaussian state $\psi_0(z)=\sum c_n\,\chi_n(z)$. They are given by an analytical expression when the probability of presence at negative altitudes is negligible, which is a very good approximation for numbers chosen for our calculations 
\begin{eqnarray}
&&c_n= \dfrac{(8\pi)^\frac{1}{4}}{\Ai'(-\lambda_n)} 
\,\sqrt{\frac{\sigma_z}{\ell_g}}
\Ai \left( \frac{z_0}{\ell_g} - \lambda_n 
+ \frac{\imath\,v_0 t_g}{\ell_g}\frac{\sigma_z^2}{\ell_g^2}  + \frac{\sigma_z^4}{\ell_g^4}  \right)   \nonumber \\
&&\;\times\exp\left(  \frac{\sigma_z^2}{\ell_g^2} 
\left( \frac{z_0}{\ell_g}-\lambda_n  
+\frac{\imath\, v_0 t_g}{\ell_g} \frac{\sigma_z^2}{\ell_g^2}
-\frac{v_0^2t_g^2}{4\ell_g^2}+ \frac{2\sigma_z^4}{3\ell_g^4}\right)\right) \,, \nonumber\\
&&\ell_g =\sqrt[3]{\frac{\hbar^2}{2gm^2}}
\quad,\quad 
e_g=m\,g\,\ell_g
\quad,\quad
t_g=\frac{\hbar}{e_g}   \,.
\label{eq:coefficientscn}
\end{eqnarray}
Relations have been written with quantities reduced to the natural units ($\ell_g$, $e_g$, $t_g$) for the GQS, and $\lambda_n$ is the absolute value of the $n-$th zero of the $\Ai-$function (energies given by $E_n=e_g\,\lambda_n$).

The second phase of the evolution corresponds to right-hand side on Figure \ref{fig:onebouncescheme}. It starts at the end of the mirror ($Vt=d$) and lasts till the detection plate ($Vt=D$). It is described by the quantum propagator which is known for quantum free fall in a constant gravity field and has the following form in momentum representation
\begin{equation}
\begin{aligned}
\widetilde{\psi}_d(p) &= \int {\psi}_d(z) \,
\exp\left(-\frac{\imath\,p\,z}{\hbar}\right) \, \frac{\md z}{\sqrt{2\pi\hbar}} 
~, \\
\widetilde{\psi}_D(p) &= \widetilde{\psi}_d(p+mgT) \\
\times& 
\exp\left(\frac{-\imath\,T}{\hbar}
 \left(\frac{p^2}{2m} +\frac{gpT}{2} + \frac{mg^2T^2}{6} \right) \right)  
~, \\
\psi_D(z) &=  \int \widetilde{\psi}_D(p) \,
\exp\left(\frac{\imath\,p\,z}{\hbar}\right) \, \frac{\md p}{\sqrt{2\pi\hbar}} ~,  \\
  &T=\frac{D-d}{V}  ~.  
\end{aligned}
\label{eq:propagator}
\end{equation} 

\begin{figure*}[t]
\begin{center}    
    \includegraphics[width=1\linewidth]{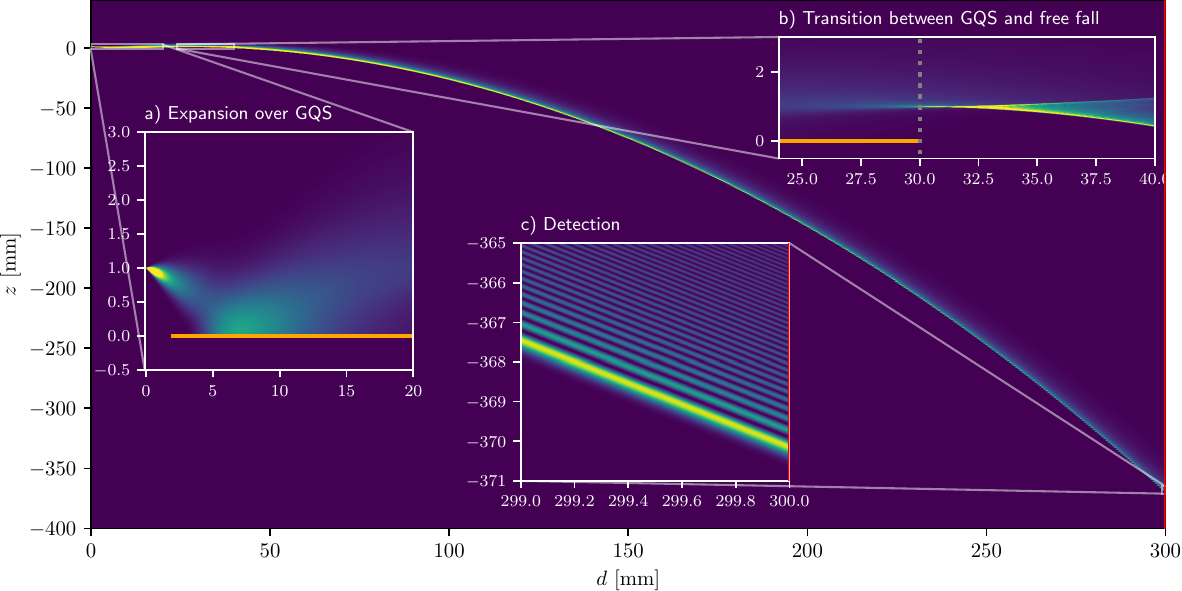}
\caption{Representation of the evolution of the wave-function from the source to the detection. The square modulus $\vert\psi(z)\vert^2$ of the wave-function is represented as a function of $x=Vt$ (for a velocity $V_0=\SI{1}{\metre\per\second}$). The main plot shows the mean motion from the source to the detection plate. Details are shown in the zooms devoted to zones of particular interest: a) evolution of the wave-function over the mirror (described by eq.\ref{eq:discrete_decomposition}); b) transition to the second phase corresponding to free fall (described by eq.\ref{eq:propagator}); c) fully revealed interference fringes after a long phase of free fall after the mirror. }
\label{fig:evolutionwavefunction}
\end{center}
\end{figure*}

At the detection plate, the measured position distribution of $\Hbar$ atom is given by the altitude-dependent density of probability $\vert\psi_{D}(z)\vert^2$. The choice of a vertical plate is made for simplifying the presentation of results. Identical results would be obtained for another orientation of the plate by adding details on the final stage of the evolution. In GBAR, the free fall chamber will be equipped with detectors installed on all surfaces, vertical and horizontal, and a dedicated analysis will have to be adapted to this geometrical configuration. 

Numbers for the plots shown below are $d=30$~mm for the horizontal distance from the source to the end of mirror and $D=300$~mm from the source to the detection plate. These numbers match the planned geometry for the GBAR experiment. We repeat that horizontal velocity $V$ is measured for each detection event by the time-of-flight, and suppose that, for all values of $V$ in the horizontal velocity distribution, atoms undergo a single bounce with a probability close to unity. 

The results of calculations of evolution of the quantum state are gathered on Figure \ref{fig:evolutionwavefunction} where the altitude-dependent  density $\vert\psi_t(z)\vert^2$ is presented as a function of run distance $x=Vt$.
The main figure represents the evolution of the density from the source to the bounce and then till the detector. 
Three zooms are shown on zones of particular interest. 

The first zone (a) shows the beginning of the evolution from the source till the middle of mirror, which includes the bounce. As the mean initial altitude $z_0$ is very large compared to the scale $\ell_g\sim6~\mu$m of GQS, we have made the numerical decomposition of the wave-function over a large number of GQS, in order to have an accurate description of evolution during this phase. With our parameters, we use  12 000 states, a number above which $c_n$ becomes negligible. In order to accommodate such a large number of states, the calculation algorithm was carefully designed to use a discrete convolution product in the calculation of $\psi_d$ and fast Fourier transforms in the calculation of free fall. Details are presented in the supplementary material.

We have chosen the dimensions of the mirror so that all atoms bounce once on it, with the mirror plate ending at 30~mm from the source. This dimension is not critical but it fits our purpose of having nearly all atoms having one bounce. We have also drawn a mirror plate beginning at 2~mm of the source, which does not change the idea but would make easier the insertion of the mirror in the planned free fall chamber for GBAR. 

The second zone (b) shows the transition from the first phase to the second one around the size of the mirror. 
After this point, we see interferences beginning to appear in the probability distribution.
The third zone (c) shows well-formed interference fringes fully revealed after a free fall during a time longer than that of the two first zones. These fringes depend on the free fall acceleration, which will lead to the accuracy of the measurement of ${\gbar}$ discussed in the following. 

One emphasizes at this point that the interferences are not produced by the superposition of two waves having followed widely separated classical trajectories, as would be the case for usual matter-wave gravimeters. In other words, there is no beam splitting in this new concept of interferometer and all quantum paths originate from a unique cell in phase space corresponding to the Heisenberg-limited initial  wave-packet.  

\begin{figure*}[t]
\begin{center}    
    \includegraphics[width=\linewidth]{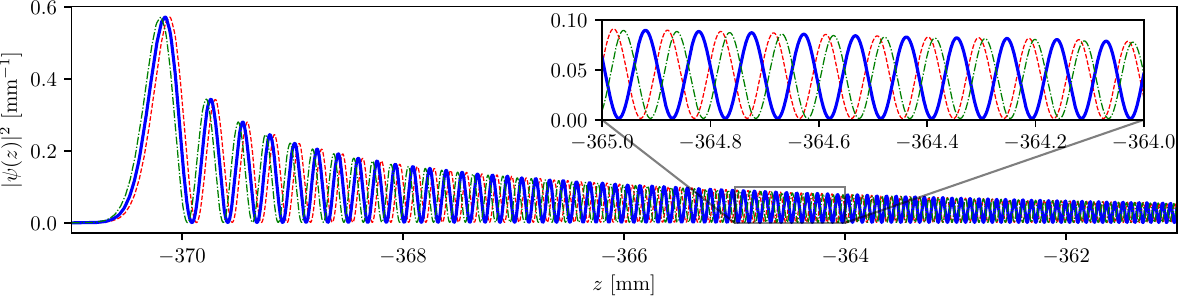}
\caption{Three curves for the probability of detection at altitude $z$ calculated for slightly different values of ${\gbar}$. The solid (blue) curve corresponds to the standard gravity acceleration, while the dashed (red, shifted towards the right) and dotted (green, shifted towards the left) curves correspond respectively to values decreased and increased by a relative variations of $10^{-4}$.  The zoom shows that fringes are visible on a large range of values of $z$. (colors online) }
\label{fig:signal}
\end{center}
\end{figure*}

At the end of this discussion of evolution, we get the altitude-dependent probability density at detector, $\vert\psi_D(z)\vert^2$ shown on Figure \ref{fig:signal}. The full blue curve is calculated for the case ${\gbar}=g_0$ where $g_0=9.81$~ms$^{-2}$ is the standard gravity acceleration at Earth surface. It shows neat interference fringes with an extremely good contrast (the curve goes to zero repeatedly when altitude varies).
The other curves correspond to slightly different values of  ${\gbar}$ with relative variations of $10^{-4}$. 
These two curves are modified essentially through a shift towards the right and left sides respectively for lower and higher values of ${\gbar}$. 
These shifts are highly visible for the slight changes considered in the calculations, and this already shows that the accuracy should be much better than $10^{-4}$, even with limited samples of detected events. 

The expected accuracy is quantitatively evaluated now by using the simulation technique proposed in \cite{Crepin2019pra}. 
In a first step, data mimicking the results of a forthcoming experiment are produced by randomly drawing $N$ events from the curve calculated for the standard value $g_0$ of the free fall acceleration.
These points are then seen as a sample of $N$ experimental detection events and a data analysis is simulated with a maximum likelihood  method.
Each drawing of a sample of $N$ data points gives an estimator $\widehat{g}$
defined as the argument maximizing the log-likelihood function. 
This numerical simulation is repeated $M$ times and a histogram of the obtained estimators $\widehat{g}$ is finally drawn.

\begin{figure}[b!]
\begin{center}
    \includegraphics[width=1\linewidth]{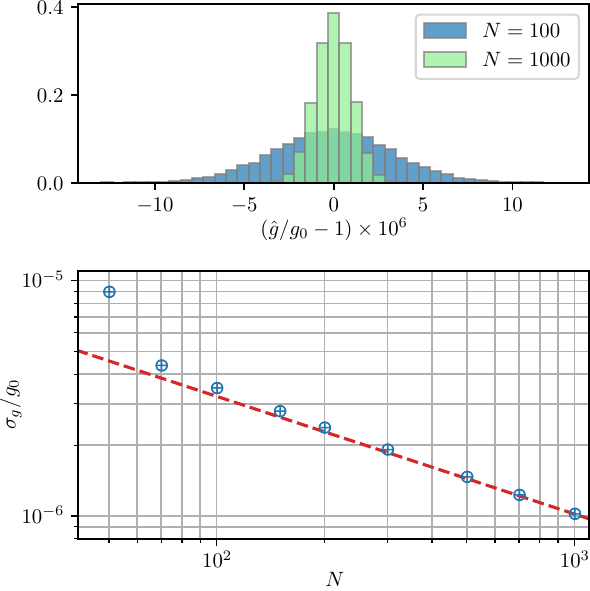}
\caption{Top plot : Histograms of the estimators $\hat{g}$ for a large number of simulations performed each with $N$ randomly drawn detection events. The broader blue histogram corresponds to $N=100$ and the narrower green one to $N=1000$. 
Bottom plot : Expected relative accuracy $\sigma_{{\gbar}}/g_0$ for the measurement of ${\gbar}$ calculated from the dispersion on histograms, and represented as blue dots depending on the number $N$. For large enough values of $N$, the dots are aligned on the dashed red line showing the Cramer-Rao law. (colors online)  }
\label{fig:histogramandprecision}
\end{center}    
\end{figure}

We show on the top part of Figure \ref{fig:histogramandprecision} the normalized histogram calculated for a large number $M=40000$ of repeated simulations, with the numbers $N=100$ and $N=1000$ of data points in each experiment.
For a large enough data sample size $N$, the histogram tends to a Gaussian distribution and the expected experimental accuracy can be predicted as the dispersion $\sigma_g$ of the distribution of estimators provided the uncertainties are dominated by statistical data sampling.
For the number $N=1000$ corresponding to the planned number of detection events in GBAR, we find an expected relative accuracy $\sigma_g/g_0$ of $\num{1.0e-6}$ which is better by roughly 4 orders of magnitude than for the classical timing experiment, with the same number of detection events.
It is even better than what was found in \cite{Crepin2019pra} for a quantum interference with many bounces although we have considered here a single-bounce scheme. 

Finally, we plot the variation of the expected relative accuracy $\sigma_{\gbar}/g_0$ versus the sample size $N$ on the bottom part of  Figure \ref{fig:histogramandprecision}.
The blue dots are the values obtained from our simulation procedure for values of $N$ ranging to $N=1000$ to the small sample size $N=50$. For large enough values of $N$, the expected accuracies vary according to the Cramer-Rao law \cite{Frechet1941,Cramer1946,Refregier2004}, shown as the dashed red line on the figure. 
The latter is fixed by the Fisher information $\mathcal{I_F}$ which can be calculated from the dependence on $g$ of the modulus of the wave-function 
\begin{equation}
\frac{\sigma_\mathrm{CR}}{g_0} = \frac{1}{\sqrt{N\,\mathcal{I_F}}} \quad,\quad
\mathcal{I_F} = 4 g_0^2 \int \left(\partial_g \vert\psi_g(z)\vert\right)^2 \,\md z  \,.
\label{eq:histogramandprecision}
\end{equation}
This means that the statistical efficiency of the measurement is good.
For the values of parameters given above, we numerically estimate the Fisher information to $\sim\num{1.e9}$, which for $N=1000$ gives an expected relative precision $\sim\num{1 e-6}$.
 For lower values of $N$, the simulated relative accuracy $\sigma_{\gbar}/g_0$ is as expected above the Cramer-Rao law. For example, the simulated value is a factor 2 larger than the Cramer-Rao bound for $N=50$. A measurement with such an under-sampled number of events goes out of the domain of applicability of common statistical techniques based on Gaussian laws. We however emphasize that adapted techniques would allow one to approach relative accuracies of the order of $\,10^{-5}$ which would still be better than classical experiment by 3 orders of magnitude.  
 
The agreement of the results of simulation techniques with the Cramer-Rao law  is a good hint of robustness of simulations (for $N>100$). Calculations based on the Fisher information are easier to perform and they can be used to explore variation of accuracy with parameters. We have used the method to choose the parameter $v_0$ once $z_0$ and $\sigma_z$ were chosen from considerations related to  GBAR experiment. It is also easy to prove in this way that the expected accuracy would be improved by increasing the free fall height before the mirror ($z_0$ larger) as well as after the mirror ($D$ larger), in full conformity with classical expectations for a free fall experiment. The free fall height after the bounce is limited by the size of free fall chamber in the experiment, and our choice above matches the plans for the GBAR experiment.

In order to discuss the effect of a variation of $z_0$, we have to be more precise on the probability of quantum reflection, which depends on the incidence energy and the specific material constituting the mirror plate.
Non reflected $\Hbar$ atoms are annihilated on the mirror and lost for the quantum interference signal. 
The probability of annihilation was very small for the low energies considered in \cite{Crepin2019pra,Rousselle2022epjd} but its importance was amplified by the large number of bounces needed to get a good accuracy. For the new concept proposed here, incidence energy is higher on the mirror, and quantum reflection probability is farther from 1, but there is only one bounce. Numbers given above for $z_0$ and $v_0$ lead to a reflection probability of $\sim 87\%$ on a bulk silica plate \cite{Dufour2013}, so that the number of detected events is reduced by approximately $\sim13\%$, which has to be taken into account in the accuracy analysis presented above. 

These numbers are improved for quantum reflection on a thick film of liquid  $^4$He \cite{Crepin2017epl} with reflection probability $\sim 97.5\%$ and loss by annihilation $\sim2.5\%$, which allows to consider larger values of $z_0$.  One notes that atoms may be quantum reflected on the detection plate instead of being annihilated there, and this loss in the number of detection events has also to be taken into account. This problem is already present for the classical GBAR experiment and all relevant numbers can be found in \cite{Dufour2013}.

As quantum reflection is not perfect, it depends on energy and this has to be included in the data analysis to avoid possible systematic errors when the experiment will be performed. The treatment will be based on the calculations of quantum reflection amplitude resulting from Casimir-Polder interaction, for example on a silica plate \cite{Crepin2019epjd}. With the single-bounce interference technique, this will be done more easily than for a many-bounces experiment, so that the measurement of ${\gbar}$ will be made more accurate and reliable than with previous proposals. 

The spatial resolution on the detection plate will also be a critical argument for an experimental implementation of the new scheme. It is indeed important to collect information on a large enough fraction of fringes in the interference pattern on Figure \ref{fig:signal} to reach the accuracy discussed above. We have simulated the behavior of the resolution of the detection and observe that with a resolution below \SI{4}{\micro\meter}, the uncertainty is not degraded. It has to be noted at this point that antimatter detectors with sub-micrometer resolution have recently been validated and will be further developed for future antihydrogen gravity measurements \cite{Berghold2025sal}. 
It is clear that a lot of work has yet to be done to estimate the noise sources not discussed in this letter, and to analyze the systematic effects affecting this measurement.

In this letter, we have proposed and studied a new concept of quantum gravimeter which differs from the known configurations for matter-wave gravimeters, where two waves interfere after beam splitters and propagations along separated classical trajectories. 
We have estimated the expected
accuracy of the new concept for the GBAR experiment, with an improvement of about 4 orders of magnitude for the measurement of free fall acceleration of $\Hbar$ atoms, using the same resource in terms of number of atoms and initial state. 
The concept is not restricted to this application and may open new ways of investigating gravitational properties of rare or exotic species, in particular when the data sample size for detection events or the time available for measurement are limited for intrinsic physical reasons.

\section*{Acknowledgments}
We thank our colleagues in  GBAR https://gbar.web.cern.ch/ and 
  GRASIAN  https://grasian.eu/ collaborations for
 insightful discussions, in particular S. Baessler, C. Blondel, 
 P. P. Blumer, C.  Christen, P.-P. Crepin, P. Crivelli, P. Debu, 
 A. Douillet, C. Drag, N. Garroum, R. Guérout, L. Hilico, 
 P. Indelicato, G. Janka, J.-P. Karr, S. Guellati-Khelifa, L. Liszkay, 
 B. Mansoulié, V. V. Nesvizhevsky, F. Nez, N. Paul, P. Pérez, 
 J. Pioquinto, C. Regenfus, O. Rousselle, F. Schmidt-Kaler, 
 K. Schreiner, A. Yu. Voronin, S. Wolf, P. Yzombard. 
 This work was supported by the Programme National GRAM of CNRS/INSU with INP and IN2P3 co-funded by CNES, and by 
 Agence Nationale pour la Recherche, Photoplus project 
 Nr. ANR 21-CE30-0047-01. Joachim Guyomard was supported by QuantEdu-France (ANR-22-CMAS-0001) in the framework of France 2030.

\bibliographystyle{elsarticle-num} 
\bibliography{bibliography}



\end{document}


\begin{frontmatter}


\title{Supplementary Material : Numerical methods to compute Gravitational Quantum States}

\affiliation[LKB]{organization={Laboratoire Kastler Brossel,Sorbonne Universit{\'{e}}, CNRS, ENS-Universit{\'{e}} PSL, Coll{\`{e}}ge de France},
            addressline={4 place Jussieu},
            city={Paris},
            postcode={75005},
            state={},
            country={France}}

\author[LKB]{Joachim Guyomard} 
\author[LKB]{ Pierre Clad\'e}
\author[LKB]{Serge Reynaud}

\begin{abstract}
We present in this document details on the methods used to calculate numerically the wave-function $\psi_D(z)$ at the detection plate. 
\end{abstract}

\end{frontmatter}


\def\arr#1{\bm{\mathrm{#1}}}
\def\ind#1{_{#1}}
\def\arrind#1#2{\arr{#1}\ind{#2}}
\def\arrA#1{\bm{#1}}
\def\arrAind#1#2{\arrA{#1}\ind{#2}}


The calculation relies on two steps, the first one consisting in calculating the wave-function $\psi_{d}(z)$ at the output of the mirror using Equation 1 and the second one in calculating the wave-function $\psi_{D}(z)$ after free fall (Equation 3). 

In this document, we  use bold symbols for numerical arrays. 

\section{Evaluation of $\psi_{d}$}
Replacing $\chi_n(z)$ by its definition in Equation 1, we obtain : 
\begin{equation}
\begin{aligned}
& \psi_{d}(z)=\phi(z)\Theta(z)\quad , \quad \phi(z)=\sum_n b_n\,\Ai(z/l_g - \lambda_n)  \\
& b_n = \frac{c_n e^{ -i \frac{E_n\, d}{V\hbar}}}{\sqrt{l_g}\Ai^\prime(-\lambda_n)}
\end{aligned}
\label{eq:discrete_decomposition_bis}
\end{equation}

For the numerical calculation, we sample the positions with a separation $\Delta_z$ between points. We write $\arr{z}$ the array of size $n_z$ defined by $\arrind{z}{j} = z_\mathrm{min} + j \Delta_z$ and call $\arrA{\upphi}\ind{j} = \phi(z_j)$. For each value $\lambda_n$ we define an integer $k_n$ such that $k_n\le\lambda_n\ell_g/\Delta_z<k_{n}+1$  and $\tau_n = \lambda_n\ell_g/\Delta_z - k_n \in[0, 1[$ and get
\begin{equation}
\arrAind{\upphi}{j} = \sum_{n=1}^{n_{GQS}} b_n\,\Ai((\arr{z}\ind{j} - k_n\Delta_z - \tau_n\Delta_z)/l_g)\, ,
\label{eq:discrete_decomposition_ter}
\end{equation}
where $n_{GQS}$ is the maximum number of quantum gravity states over which the sum is made.

In order to improve the efficiency of the calculation, we would like to implement this equation as a discrete convolution product. This is not possible directly because the term in $\tau_n\Delta_z$ does not correspond to an integer shift of the index.

To circumvent this problem, we use an interpolation, precisely a cubic Hermite spline interpolation. In this case, the term $n$ depends on $\arr{A}\ind{i-k_n}$ and $\arr{A}\ind{i-k_n-1}$ as well as $\arr{A^\prime}\ind{i-k_n}$ and $\arr{A^\prime}\ind{i-k_n-1}$, where $\arr{A}\ind{i} = \operatorname{Ai}(z_i)$ and $\arr{A^\prime}\ind{i} = \operatorname{Ai}^\prime(z_i)$. We obtain :  
\begin{equation}
\begin{aligned}
 \arrA{\upphi} = &\left(\sum_n b_n \arrA{\updelta}^{k_n}h_{00}(\tau_n) +  b_n\arrA{\updelta}^{k_n+1}h_{01}(\tau_n)  \right) \ast\arr{A} \\ &- \left(\sum_n b_n \Delta_z\arrA{\updelta}^{k_n}h_{10}(\tau_n) +  b_n\Delta_z\arrA{\updelta}^{k_n+1}h_{11}(\tau_n)  \right) \ast\arr{A^\prime}
\end{aligned}
\end{equation}
where
\begin{equation}
\begin{aligned}
    & h_{00}(t) = 2t^3-3t^2+1 \,;\, h_{10}(t) = t^3-2t^2+t\,; \,\\
& h_{01}(t) = -2t^3+3t^2 \,;\,
h_{11}(t) = t^3-t^2   \,,\\
\end{aligned}
\end{equation} the array $\arrA{\updelta}^{k}$ is defined by: 
\begin{equation}
    \arrA{\updelta}^{k}_i=\begin{cases} 1 & \text{for}\: i = k \\0 & \text{otherwise} \end{cases}
\end{equation}  and $\ast$ represents the convolution product : 
\begin{equation}(\arr{a}\ast\arr{b})\ind{i} = \sum_j \arr{a}\ind{j}\arr{b}\ind{i-j} \quad .
\end{equation} 

This implementation represents several advantages : \textit{i)} the Airy function in now evaluated on a single set of $n_z$ points instead of the $n_z\times n_{GQS}$, \textit{ii)} similarly the left handside of each convolution product can be calculated using $O\left(n_{GQS}\right)$ operations and \textit{iii)} when implemented using fast Fourier transform (FFT), the complexity of this convolution products scales as $O\left(n_z\log{n_z}\right)$\cite{numerical_recipes_2007}

\section{Evaluation of $\psi_{D}$}
\label{sec:evalutoin_psi_D}
To evaluate Equation 3 and take into account the momentum shift $ mgT$ due to gravity, we define an intermediate function $a(p) = \widetilde{\psi}_D(p - mgT)$ and obtain : 
\begin{equation}
\begin{aligned}
\psi_D(z) &=  \left(\int a(p) \,
\exp\left(\frac{\imath\,p\,z}{\hbar}\right) \, \frac{\md q}{\sqrt{2\pi\hbar}}\right) \exp\left(\frac{-\imath\,mgTz}{\hbar}\right)~.  
\end{aligned}
\label{eq:propagator_bis}
\end{equation} 

For an array $\arr{x}$ of size $N$ we write its discrete Fourier transform 
\begin{equation}
    \mathcal{F}(\arr{x})\ind{j} = \sum_k \arr{x}\ind{k} \exp\left(-2\imath \pi \frac{kj}{N}\right)
    ~,
\end{equation}
and its inverse transform 
\begin{equation}
    \mathcal{F}^{-1}(\arrA{\tilde{x}})\ind{j} = \frac{1}{N}\sum_k \arrA{\tilde{x}}\ind{k} \exp\left(2\imath \pi \frac{kj}{N}\right)
\end{equation}
They are evaluated by fast Fourier transform algorithm (FFT) with a complexity $O\left(N \log{N}\right)$ \cite{numerical_recipes_2007}.

When $\arrA{\uppsi}$ represents a wave-function in position, $ \mathcal{F}(\arrA{\uppsi})$ represents the  wave-function in momentum space with index $j$ corresponding to momentum $\arr{p}\ind{j}$ given by :
\begin{equation} \arr{p}\ind{j}= 
\begin{cases} \frac{2\pi \hbar j}{n_z\Delta_z } & \text{for}\: j<n_z/2 \\\frac{2\pi\hbar (j-n_z)}{n_z\Delta_z} & \text{for}\: j\ge n_z/2 \end{cases}
\end{equation}

The free fall is thus described by the following three steps 
\begin{equation}
\begin{aligned}
& \arrA{\widetilde{\uppsi}^d} = \mathcal{F}(\arr{{\uppsi}_d}) \\
    &\arr{a}\ind{j} = \arrA{\widetilde{\uppsi}^d}\ind{j}
\exp\left(\frac{-\imath\,T}{\hbar}
 \left(\frac{(\arr{p}\ind{j}-mgT)^2}{2m} +\frac{g(\arr{p}\ind{j}-mgT)T}{2} + \frac{mg^2T^2}{6} \right) \right) \\
& \arrA{\uppsi^D}\ind{j} = \exp\left(\frac{-\imath\,mgT\,\arr{z}\ind{j}}{\hbar}\right)  \mathcal{F}^{-1}(\arr{a})\ind{j}
\end{aligned}
\end{equation}

As all operation are linear, it is not necessary to take into account the normalization constants in the definition of the Fourier transform and its inverse. 

\section{Choice of the parameters}

The coefficient $c_n$ represents the amplitude of the component of $\psi$ at energy $E_n = \lambda_n e_g$. The energy increases with $n$ and, for large value of $n$, $c_n$ becomes negligible. The maximal value of $n$ denoted $n_\mathrm{GQS}$ is estimated numerically by checking that $\sum_{n=0}^{n_\mathrm{GQS}} |c_n|^2 \simeq 1$. For our parameters, we have $\left|\sum_{n=0}^{n_\mathrm{GQS}} |c_n|^2 -1 \right| < \num{1E-4}$ for $n_\mathrm{GQS}=12\,000$. 

Once $n_\mathrm{GQS}$ is chosen, one fixes a value of $\Delta_z$ small enough to fulfill the Nyquist criteria. The maximal momentum is given by $\pi\hbar/\Delta_z$, leading to the criteria $\Delta_z \ll {\pi\hbar}/{\sqrt{2mE_{n_\mathrm{GQS}}}}  = {\pi\ell_g}/{\sqrt{\lambda_{n_\mathrm{GQS}}}}$

For our simulation, $n_\mathrm{GQS}=12\,000$, ${\pi\ell_g}/{\sqrt{\lambda_{n_\mathrm{GQS}}}}=\SI{4.5E-7}{\meter}$
and we have chosen $\Delta_z=\SI{9.8e-8}{\meter}$. We have numerically tested the convergence of the algorithm for this value.

The Airy function $\operatorname{Ai}(x)$ decreases very quickly for positive values of $x$ and becomes completely negligible for $x>x_\mathrm{max}=10$, therefore, the maximum height above which $\psi$ will be negligible is given by $z_\mathrm{max} \simeq (\lambda_{n_\mathrm{GQS}} + x_\mathrm{max})l_g$. Due to the convolution algorithm, one need to choose, $z_\mathrm{min} \le -z_\mathrm{max}$. 

In order to optimize the calculation, we choose $z_\mathrm{min} = -z_\mathrm{max}$ for the first part. For the second part, we extend the grid and choose a value of $z_\mathrm{min}$ small enough to take into account the free fall. In both cases, the number of points is chosen to be a power of 2, which is optimal for the convolution product and the FFT algorithm.

\bibliographystyle{elsarticle-num} 
\bibliography{bibliography_supp}